\documentclass[aps,prd,superscriptaddress,nofootinbib]{revtex4}%
\usepackage{graphicx}
\usepackage{bm,latexsym,amsmath,amssymb,amsfonts,mathrsfs}
\usepackage{color}
\input{colordvi.tex}

\usepackage[dvipdfmx]{hyperref}
\usepackage{url}
\usepackage{ulem}
\usepackage{hyperref}
\hypersetup{
 colorlinks=true,
  citecolor=cyan,
 }
\allowdisplaybreaks[1]

\usepackage[hang,small,bf]{caption}
\usepackage[margin=0pt]{subcaption}
\newcommand*{\D}{{\rm d}}

\begin{document}

\title{Towards testing the general bounce cosmology with the CMB B-mode auto-bispectrum}
\author{Shingo~Akama}
\email[Email: ]{shingo.akama"at"uj.edu.pl}
\affiliation{Faculty of Physics, Astronomy and Applied Computer Science, Jagiellonian University, 30-348 Krakow, Poland
}
\author{Giorgio~Orlando}
\email[Email: ]{giorgio.orlando"at"uj.edu.pl}
\affiliation{Faculty of Physics, Astronomy and Applied Computer Science, Jagiellonian University, 30-348 Krakow, Poland
}
\author{Paola~C.~M.~Delgado}
\email[Email: ]{paola.moreira.delgado"at"doctoral.uj.edu.pl}
\affiliation{Faculty of Physics, Astronomy and Applied Computer Science, Jagiellonian University, 30-348 Krakow, Poland
}

\begin{abstract}
It has been shown that a three-point correlation function of tensor perturbations from a bounce model in general relativity with a minimally-coupled scalar field is highly suppressed, and the resultant three-point function of cosmic microwave background (CMB) B-mode polarizations is too small to be detected by CMB experiments. On the other hand, bounce models in a more general class with a non-minimal derivative coupling between a scalar field and gravity can predict the three-point correlation function of the tensor perturbations without any suppression, the amplitude of which is allowed to be much larger than that in general relativity. In this paper, we evaluate the three-point function of the B-mode polarizations from the general bounce cosmology with the non-minimal coupling and show that a signal-to-noise ratio of the B-mode auto-bispectrum in the general class can reach unity for $\ell_{\rm max}\geq9$ and increase up to $5.39$ for $\ell_{\rm max}=100$ in the full-sky case. We also discuss the possibility to test the general class of bounce models by upcoming CMB experiments.

\end{abstract}

\maketitle


\section{Introduction}
Inflation~\cite{Guth:1980zm,Sato:1981qmu,Linde:1983gd} is the most conventional scenario of the early universe, whereas, motivated by the initial singularity problem~\cite{Borde:1996pt} in that scenario, non-singular alternatives have also been considered (see, e.g., Refs.~\cite{Creminelli:2010ba,Battefeld:2014uga,Brandenberger:2016vhg}). The study of the non-singular alternatives is not only interesting for exploring theoretical new possibilities but also important for testing inflation itself by comparing the observational predictions from different scenarios. For these reasons, it is worth investigating observational signatures in the non-singular alternatives as well as inflation. 

In the present paper, we consider a bouncing cosmology~\cite{Battefeld:2014uga,Brandenberger:2016vhg} as the non-singular alternative scenario. Recent studies have found a model space of the bouncing cosmology where the observed cosmic microwave background (CMB) fluctuations are predicted consistently~\cite{Akama:2019qeh}. The CMB temperature fluctuation and E-mode polarization originate from both a primordial density fluctuation and primordial gravitational waves, while the B-mode polarization arises solely from the latter~\cite{Seljak:1996gy}. Statistical properties of the primordial gravitational waves have been less constrained compared to those of the primordial density fluctuation. However, both fluctuations carry rich information on the early universe physics, and thus future detections of the B-mode polarization by e.g., LiteBIRD~\cite{LiteBIRD:2022cnt}, CMB-S4~\cite{Abazajian:2019eic}, Simons Observatory~\cite{SimonsObservatory:2018koc}, SPIDER~\cite{SPIDER:2021ncy}, BICEP Array~\cite{Moncelsi:2020ppj}, and PICO~\cite{NASAPICO:2019thw}, among others, are expected to open a new window to a more detailed description of the early universe.

In general, two-point statistics of the gravitational waves from inflation can be mimicked by non-singular alternatives (see, e.g., Ref.~\cite{Wands:1998yp}), and hence different scenarios can predict the same two-point statistics of the B-mode polarization. On the other hand, the degeneracy of the predictions at the power spectrum level can be resolved at the primordial bispectrum level~\cite{Kothari:2019yyw,Akama:2019qeh}. Therefore, the three-point statistics of the B-mode polarization originating from the primordial tensor bispectrum are important for discriminating the different scenarios.

So far, the three-point statistics of the B-mode polarization from the bouncing cosmology has been studied only in the context of general relativity with a minimally-coupled scalar field~\cite{Kothari:2019yyw}, where it has been found that a signal-to-noise ratio of the B-mode auto-bispectrum is too small to be detected. 
On the other hand, those from inflation have been widely studied. In particular, a recent study has shown that inflation in the context of general relativity predicts a signal-to-noise ratio much smaller than unity~\cite{Tahara:2017wud}. It has been found that for extended models of inflation in beyond general relativity, the three-point function of the B-mode polarization can potentially be detected by CMB experiments~\cite{Tahara:2017wud,Shiraishi:2019yux} (see also Refs.~\cite{Shiraishi:2011dh,Shiraishi:2012rm,Shiraishi:2013kxa} for a sizable B-mode bispectrum originating from external sources). Thus, as with the extended models of inflation, it is interesting to see whether observable signatures of the three-point statistics are predicted in an extended framework of the bouncing cosmology, e.g. modified gravity.

In our paper, we use the Horndeski theory which is the most general scalar-tensor theory giving the second-order field equations~\cite{Horndeski:1974wa,Deffayet:2011gz,Kobayashi:2011nu} (see Ref.~\cite{Kobayashi:2019hrl} for a review). An extended framework of the bouncing cosmology with a power-law contracting phase has been constructed and the three-point function of tensor perturbations has been computed in this theory~\cite{Akama:2019qeh} (see also Ref.~\cite{Nandi:2019xag} for a similar analysis in a subclass of the Horndeski theory). Besides a cubic operator of the tensor perturbations present in general relativity, an additional one is induced by the so-called $G_5$ term in the Horndeski Lagrangian~\cite{Gao:2011vs}. In particular, for the bouncing cosmology with a minimally-coupled scalar field, the three-point function of the tensor perturbations originating from the cubic operator in general relativity is negligibly small~\cite{Chowdhury:2015cma,Akama:2019qeh}, which results in unobservable three-point statistics of the B-mode polarization~\cite{Kothari:2019yyw}. The three-point function originating from the additional cubic operator does not have such a negligibly small amplitude~\cite{Akama:2019qeh}, but its impact on the B-mode polarization has not been studied so far. In the present paper, by using the extended framework of the bouncing cosmology, we calculate the signal-to-noise ratio of the B-mode bispectrum originating from the new cubic interaction term and show that the three-point statistics can potentially be tested by upcoming CMB experiments. 

This paper is outlined as follows. In the next section, we briefly review the framework of general bounce cosmology including both primordial power spectum and bispectrum of tensor perturbations. In Sec.~\ref{Sec: B-mode}, we first show an analytical expression of a B-mode auto-bispectrum and then compute it numerically. In the same section, we also compute a signal-to-noise ratio of the B-mode bispectrum and evaluate its upper bound by taking into account a condition for linear perturbation theory to be valid. Our conclusions are given in Sec.~\ref{Sec: conclusion}.

\section{Setup}\label{Sec: Setup}
Throughout the present paper, we work in a spatially-flat Friedmann-Lema\^{i}tre-Robertson-Walker (FLRW) metric which is of the form,
\begin{align}
\D s^2&=-\D t^2+a^2(t)\delta_{ij}\D x^i \D x^j\notag\\
&=a^2(\eta)(-\D\eta^2+\delta_{ij}\D x^i \D x^j) \, , \label{eq: bgmetric}
\end{align}
where $a$ is the scale factor and $\eta$ stands for the conformal time defined by $\D\eta:=\D t/a$. Hereafter, a dot and a prime represent differentiation with respect to $t$ and $\eta$, respectively. For later convenience, we introduce the Hubble parameter defined by $H:=\dot a/a$. The early universe in the bouncing cosmology experiences contracting, bouncing, and expanding phases. The simplest contracting phase is described by a power-law model in which the scale factor is parametrized as
\begin{align}
a=\biggl(\frac{-t}{-t_*}\biggr)^n=\biggl(\frac{-\eta}{-\eta_*}\biggr)^{n/(1-n)} \, ,
\end{align}
where $0<n<1$, and $t_*(<0)$ and $\eta_*(<0)$ represent the cosmic and conformal time at the end of the contracting phase, respectively. As an example, one can consider the case of $n=2/3$, which is known as the matter bounce~\cite{Brandenberger:2012zb} and whose contracting phase is dominated by matter. Classical power-law contracting models have been unified in the Horndeski theory, the action of which is of the form~\cite{Horndeski:1974wa,Deffayet:2011gz,Kobayashi:2011nu}
\begin{align}
S&=\int\D^4 x\sqrt{-g}\biggl[G_2(\phi,X)-G_3(\phi,X)\Box\phi+G_4(\phi,X)R+G_{4X}[(\Box\phi)^2-(\nabla_{\mu}\nabla_{\nu}\phi)^2]\notag\\
&\quad +G_5(\phi,X)G_{\mu\nu}\nabla^{\mu}\nabla^{\nu}\phi-\frac{1}{6}G_{5X}[(\Box\phi)^3-3\Box\phi(\nabla_{\mu}\nabla_{\nu}\phi)^2+2(\nabla_{\mu}\nabla_{\nu}\phi)^3]\biggr] \, , \label{eq: Horndeski action}
\end{align}
where $G_i(\phi,X)$ are arbitrary functions of $\phi$ and its canonical kinetic term $X:=-g^{\mu\nu}\nabla_{\mu}\phi\nabla_{\nu}\phi/2$. In the present paper, we define $G_X:=\partial_X G$ and $G_\phi:=\partial_\phi G$. 

Whether the statistical properties of the perturbations during the contracting phase can have imprints from the subsequent phases or not depends on the specifications of the model considered (see, e.g., Refs.~\cite{Gao:2009wn,Quintin:2015rta}). In the present paper, we follow Ref.~\cite{Akama:2019qeh} and assume that the observational predictions are solely determined during the contracting phase, and hence we investigate the impacts of a correlation function evaluated at the end of the contracting phase on the statistics of the CMB B-mode polarization.

Under the background metric in Eq.~(\ref{eq: bgmetric}), the Friedmann and evolution equations have been obtained, respectively, as~\cite{Kobayashi:2011nu}
\begin{align}
\mathcal{E}&=\sum_{i=2}^5\mathcal{E}_i\notag\\
&=2XG_{2X}-G_2+6X\dot\phi{H}G_{3X}-2XG_{3\phi}-6H^2G_4+24H^2{X}\left(G_{4X}+XG_{4XX}\right)\notag\\
&\quad-12HX\dot\phi{G_{4\phi{X}}}-6H\dot\phi{G_{4\phi}}+2H^3X\dot\phi\left(5G_{5X}+2XG_{5XX}\right)-6H^2X\left(3G_{5\phi}+2XG_{5\phi{X}}\right)=0 \, ,\\
\mathcal{P}&=\sum_{i=2}^5\mathcal{P}_i\notag\\
&=G_2-2X(G_{3\phi}+\ddot{\phi}G_{3X})+2(3H^2+2\dot{H})G_4-12H^2XG_{4X}-4H\dot{X}G_{4X}-8\dot{H}XG_{4X}\notag\\
&\quad-8HX\dot{X}G_{4XX}+2(\ddot{\phi}+2H\dot{\phi})G_{4\phi}+4XG_{4\phi\phi}+4X(\ddot{\phi}-2H\dot{\phi})G_{4\phi{X}}-2X(2H^3\dot{\phi}+2H\dot{H}\dot{\phi}+3H^2\ddot{\phi})G_{5X}
\notag\\
&\quad-4H^2X^2\ddot{\phi}G_{5XX}+4HX(\dot{X}-HX)G_{5\phi{X}}+2\left[2(HX)^{{\boldsymbol \cdot}}+3H^2X\right]G_{5\phi}+4HX\dot{\phi}G_{5\phi\phi}=0 \, ,
\end{align}
where $\mathcal{E}_i$ and $\mathcal{P}_i$ are originating from the $G_i$ terms in the action~(\ref{eq: Horndeski action}). The general class of power-law contracting models has been studied by assuming that each term in $\mathcal{E}$ and $\mathcal{P}$ evolves in time as
\begin{align}
\mathcal{E}_i, \mathcal{P}_i\propto (-t)^{2\alpha} \, . \label{eq: scaling-E-P}
\end{align}

The tensor perturbations $h_{ij}$ around the FLRW metric are defined by
\begin{align}
\D s^2=-\D t^2+ g_{ij}\D x^i\D x^j \, ,
\end{align}
where 
\begin{align}
g_{ij}=a^2\biggl(\delta_{ij}+h_{ij}+\frac{1}{2}h_{ik}h_{kj}+\frac{1}{6}h_{ik}h_{kl}h_{lj}\cdots\biggr) \, .
\end{align}
By expanding Eq.~(\ref{eq: Horndeski action}) up to quadratic and cubic order in the perturbations, one obtains the quadratic and cubic actions as~\cite{Kobayashi:2011nu} 
\begin{align}
S^{(2)}_h&=\frac{1}{8}\int\D t\D^3 xa^3\biggl[\mathcal{G}_T\dot h_{ij}^2-\frac{\mathcal{F}_T}{a^2}(\partial_k h_{ij})^2\biggr],\\
S^{(3)}_h&=\int\D t\D^3 x a^3\biggl[\frac{\mathcal{F}_T}{4a^2}\biggl(h_{ik}h_{jl}-\frac{1}{2}h_{ij}h_{kl}\biggr)h_{ij,kl}+\frac{\mu}{12}\dot h_{ij}\dot h_{jk}\dot h_{ki}\biggr]=:-\int\D t H_{\rm int} \, , \label{eq: cubic-int}
\end{align}
where 
\begin{align}
\mathcal{G}_T&:=2[G_4-2XG_{4X}-X(H\dot\phi G_{5X}-G_{5\phi})] \, ,\\
\mathcal{F}_T&:=2[G_4-X(\ddot\phi G_{5X}+G_{5\phi})] \, ,
\end{align}
and $\mu:=\dot\phi X G_{5X}$.
Here, we have $\mathcal{G}_T, \mathcal{F}_T\sim \mathcal{E}_4/H^2, \mathcal{P}_4/H^2\propto (-t)^{2(\alpha+1)}$, which indicates that the propagation speed of the gravitational waves $c_h^2:=\mathcal{F}_T/\mathcal{G}_T$ is a constant. Additionally, one can express $\mu$ as $\mu=-(1/2)\partial\mathcal{G}_T/\partial H$, and hence we have $\mu\propto(-t)^{2\alpha+3}$ in our power-law setup.  

In Fourier space, we expand the quantized perturbations as
\begin{align}
h_{ij}(t,{\bf x})&=\int\frac{\D^3k}{(2\pi)^3}\hat h_{ij}(t,{\bf k})e^{i\bf k\cdot\bf x},\notag\\
&=\sum_{s}\int\frac{\D^3k}{(2\pi)^3}[h^{(s)}_{\bf k}(t)\hat a^{(s)}_{\bf k}e^{(s)}_{ij}({\bf k})+h^{(s)*}_{-\bf k}(t)\hat a^{(s)\dagger}_{-\bf k}e^{(s)*}_{ij}(-{\bf k})]e^{i\bf k\cdot\bf x} \, ,
\end{align}
where the creation and annihilation operators satisfy the following commutation relations:
\begin{align}
[\hat a^{(s)}_{\bf k},\hat a^{(s')\dagger}_{\bf k'}]&=(2\pi)^3\delta_{ss'}\delta({\bf k}-{\bf k}') \, ,\\
{\rm others}&=0 \, ,
\end{align}
with $s=\pm2$ standing for the two helicity states of the graviton. The polarization tensor $e^{(s)}_{ij}$ satisfies the transverse and traceless conditions, $k_i e^{(s)}_{ij}=0=\delta_{ij}e^{(s)}_{ij}$. Additionally, we adopt the normalization condition as $e^{(s)}_{ij}({\bf k})e^{(s')*}_{ij}({\bf k})=2\delta_{ss'}$.

Let us express the two-point function of the tensor perturbations in terms of the power spectrum $\mathcal{P}_h$ as
\begin{align}
\langle\hat \xi^{(s)}({\bf k}) \hat\xi^{(s')*}({\bf k}')\rangle&=:(2\pi)^3\delta_{ss'}\delta({\bf k}+{\bf k}')\frac{\pi^2}{k^3}\mathcal{P}_h \, ,
\end{align}
where we introduced $\hat\xi({\bf k}):=\hat h_{ij}({\bf k})e^{(s)*}_{ij}/2$.
Given the power-law behavior of the time-dependent functions (i.e., $a, \mathcal{G}_T$ and $\mathcal{F}_T$), the equation of motion for canonically normalized tensor perturbations can be written as
\begin{align}
v^{(s)''}_{\bf k}+\biggl[c_h^2k^2-\frac{1}{\eta^2}\biggl(\nu^2-\frac{1}{4}\biggr)\biggr]v^{( s)}_{\bf k}=0 \, ,
\end{align}
where $v^{(s)}_{\bf k}:=a(\mathcal{G}_T\mathcal{F}_T)^{1/4}/2)h^{(s)}_{\bf k}$, and we defined $\nu$ by
\begin{align}
\nu:=\frac{-1-3n-2\alpha}{2(1-n)} \, .
\end{align}
By imposing the Minkowski vacuum initial condition for the perturbations as
\begin{align}
\lim_{\eta\to-\infty}v^{(s)}_{\bf k}=\frac{1}{\sqrt{2k}}e^{-ic_h k \eta} \, ,
\end{align}
one obtains the solution of the mode function as
\begin{align}
h^{(s)}_{\bf k}&=\frac{2}{a(\mathcal{G}_T\mathcal{F}_T)^{1/4}}\cdot\frac{\sqrt{\pi}}{2}\sqrt{-c_h\eta}H^{(1)}_\nu(-c_hk\eta) \, .
\end{align}
The spectral index of the tensor power spectrum, $n_t:=\D\ln\mathcal{P}_h/\D\ln k$, can be written in terms of $\nu$ as~\cite{Akama:2019qeh}
\begin{align}
n_t=3-2|\nu|.
\end{align}
Since the equation of motion for $v^{(s)}_{\bf k}$ is invariant under $\nu\to-\nu$, the same spectral index is realized for two cases. Note that a spectral index of a power spectrum of curvature perturbations $n_s$ takes the same form, i.e., $n_s-1=n_t=3-2|\nu|$~\cite{Akama:2019qeh}. 
The scale-invariant case corresponds to $|\nu|=3/2$ that is realized for $\alpha=-2 \ (\nu=3/2)$ or $\alpha=1-3n\ (\nu=-3/2)$. These two branches are related to the behavior of superhorizon modes satisfying $|c_hk_i\eta|\ll1$ since the amplitudes of those modes are proportional to $|\eta|^{\nu-|\nu|}$: the amplitudes for $\alpha=-2\ (\nu=3/2>0)$ are constant as in de Sitter universe while those grow in time in proportion to $|\eta|^{-3}$ as in a matter-dominated contracting universe. In terms of conformal transformation, the authors of Ref.~\cite{Akama:2019qeh} have shown that the dynamics of the perturbations in the former and the latter are identical to de Sitter inflation and matter-dominated contracting models, respectively (see Ref.~\cite{Wands:1998yp} for the duality of de Sitter inflation and the matter-dominated contracting universes in terms of the scale invariance of the power spectra). Since observational consequences of the perturbations in de Sitter spacetime have been well studied so far, we solely consider the case of $\alpha=1-3n$.\footnote{See Ref.~\cite{Nandi:2019xag} where the bounce model generating the tensor perturbations effectively living in de Sitter was studied in a subclass of the Horndeski theory, which corresponds to the case of $\alpha=-2$ in our paper.} Also, we do not discuss scalar perturbations in our paper, but it has been shown in Ref.~\cite{Akama:2019qeh} that the case of $\alpha=1-3n$ can realize a small scalar non-Gaussianity and a small tensor-to-scalar ratio, consistent with the observed CMB fluctuations. 

In the case of $\alpha=1-3n$, the power spectrum has been obtained as~\cite{Akama:2019qeh}
\begin{align}
\mathcal{P}_h&=\frac{2}{\pi^2}\biggl(1-\frac{1}{n}\biggr)^2\frac{H^2}{\mathcal{F}_Tc_h}\biggl|_{t=t_*} \, .
\end{align}
Note that the power spectrum grows with time since the mode function does.

We then move to the three-point function of the tensor perturbations. By employing the in-in formalism and using the interaction Hamiltonian defined in Eq.~(\ref{eq: cubic-int}), one can compute the three-point function as
\begin{align}
\langle\hat\xi^{(s_1)}({\bf k}_1)\hat\xi^{(s_2)}({\bf k}_2)\hat\xi^{(s_3)}({\bf k}_3)\rangle&=-i\int^{\eta_*}_{-\infty}\D\eta'a(\eta')\langle[\hat\xi^{(s_1)}(\eta_*,{\bf k}_1)\hat\xi^{(s_2)}(\eta_*,{\bf k}_2)\hat\xi^{(s_3)}(\eta_*,{\bf k}_3),H_{\rm int}(\eta')]\rangle,\\
&=(2\pi)^3\delta({\bf k}_1+{\bf k}_2+{\bf k}_3)(\mathcal{B}_{\rm GR}+\mathcal{B}_{\rm new}) \, ,
\end{align}
where $\mathcal{B}_{\rm GR}$ and $\mathcal{B}_{\rm new}$ stand for the bispectrum originating from the cubic operators of the form $h^2\partial^2 h$ and $\dot h^3$, respectively. For later convenience, let us express those as
\begin{align}
\mathcal{B}_{\rm GR}&:=\frac{(2\pi)^4\mathcal{P}_h^2}{k_1^2k_2^2k_3^2}\mathcal{A}_{\rm GR} \, ,\\
\mathcal{B}_{\rm new}&:=\frac{(2\pi)^4\mathcal{P}_h^2}{k_1^2k_2^2k_3^2}\mathcal{A}_{\rm new} \, .
\end{align}
In the general bounce cosmology, the explicit forms of $\mathcal{A}_{\rm GR}$ and $\mathcal{A}_{\rm new}$ have been obtained as~\cite{Akama:2019qeh}
\begin{align}
\mathcal{A}_{\rm GR}&=-\frac{1}{128}c_h^2K^2\eta_*^2\biggl(\frac{\sum_i k_i^3}{k_1k_2k_3}\biggr)S_{\rm GR} \, ,\\
\mathcal{A}_{\rm new}&=\frac{3}{16}\frac{1-n}{n}\frac{\mu H}{\mathcal{G}_T}\biggl(\frac{\sum_i k_i^3}{k_1k_2k_3}\biggr)S_{\rm new} \, ,
\end{align}
where $K:=k_1+k_2+k_3$ and
\begin{align}
S_{\rm GR}&:=\frac{1}{K^2}\biggl[k_{3k}k_{3l}e^{(s_1)*}_{ki}e^{(s_2)*}_{ij}e^{(s_3)*}_{jl}+\frac{1}{2}k_{2i}k_{3k}e^{(s_1)*}_{ik}e^{(s_2)*}_{jl}e^{(s_3)*}_{jl}+(2\ {\rm permutations})\biggr] \, ,\\
S_{\rm new}&:=e^{(s_1)*}_{ij}e^{(s_2)*}_{jk}e^{(s_3)*}_{ki} \, .
\end{align}
$\mathcal{A}_{\rm GR}$ involves the suppression factor $|c_hK\eta|^2\ll1$, which makes the detection of the B-mode auto-bispectrum from that term hopeless. Actually, in Ref.~\cite{Kothari:2019yyw}, a signal-to-noise ratio of a CMB B-mode auto-bispectrum has been computed from $\mathcal{A}_{\rm GR}$ in the context of matter bounce cosmology (i.e., the case of $n=2/3$), but the smallness of the three-point function of the tensor perturbations has led to the signal-to-noise ratio much smaller than unity. Furthermore, it has been shown in the context of inflation as well that the B-mode auto-bispectrum from $\mathcal{A}_{\rm GR}$ leads to a small signal-to-noise ratio so that we cannot detect the B-mode at the bispectrum level. In particular, the authors of Ref.~\cite{Tahara:2017wud} have clarified that only $\mathcal{A}_{\rm new}$ has a potential to be tested by CMB experiments depending on the value of $\mu H/\mathcal{G}_T$ within the Horndeski theory. In light of the above, in the following section, we will compute the B-mode auto-bispectrum originating solely from the new cubic operator and evaluate the signal-to-noise ratio.

\section{B-mode auto-bispectrum}\label{Sec: B-mode}

The CMB B-mode polarization, $B(\hat n)$, originating from the primordial tensor perturbations can be expanded on the CMB sky in terms of the spherical harmonics as
\begin{align}
B(\hat n)=\sum_{\ell,m}a^{\rm B}_{\ell m}Y_{\ell m}(\hat n) \, ,
\end{align}
where 
\begin{align}
a^{\rm B}_{\ell m}:=4\pi(-i)^{\ell}\int\frac{\D^3k}{(2\pi)^3}\mathcal{T}^{{\rm B}}_{\ell}(k)\sum_s\biggl(\frac{s}{2}\biggr){}_{-s}Y^*_{{\ell}m}(\hat n)\xi^{(s)}({\bf k}) \, ,
\end{align}
with the B-mode transfer function $\mathcal{T}^{{\rm B}}_{\ell}(k)$ and the spin-weighted spherical harmonics ${}_{s}Y_{{\ell}m}$. Throughout the present paper, we use the transfer function computed using the Boltzman code CAMB~\cite{CAMB:note} with the latest values of cosmological parameters obtained by Planck~\cite{Planck:2018vyg}.

Then the B-mode angular-averaged auto-bispectrum denoted by $B_{\ell_1\ell_2\ell_3}$ can be written as
\begin{align}
B_{\ell_1\ell_2\ell_3}&=\sum_{m_1,m_2,m_3}
\begin{pmatrix}
\ell_1 & \ell_2 & \ell_3 \\
m_1 & m_2 & m_3
\end{pmatrix}
\langle a_{\ell_1m_1}a_{\ell_2m_2}a_{\ell_3m_3}\rangle\notag\\
&=\sum_{s_1,s_2,s_3}B^{(s_1s_2s_3)}_{\ell_1\ell_2\ell_3} \, ,
\end{align}
where
$
\begin{pmatrix}
\ell_1 & \ell_2 & \ell_3 \\
m_1 & m_2 & m_3
\end{pmatrix}
$
is the Wigner-$3j$ symbol, and 
\begin{align}
B^{(s_1s_2s_3)}_{\ell_1\ell_2\ell_3}&=\sum_{m_1,m_2,m_3}
\begin{pmatrix}
\ell_1 & \ell_2 & \ell_3 \\
m_1 & m_2 & m_3
\end{pmatrix}
\prod_{j=1}^3\biggl[4\pi(-i)^{{\ell}_j}\int\frac{\D^3 k}{(2\pi)^3}\biggl(\frac{s_j}{2}\biggr){}_{-s_j}Y^*_{{\ell}_jm_j}(\hat n_j)\mathcal{T}_{{\ell}_j}(k_j)\biggr]\notag\\
&\quad \times\langle\hat\xi^{(s_1)}({\bf k}_1)\hat\xi^{(s_2)}({\bf k}_2)\hat\xi^{(s_3)}({\bf k}_3)\rangle \, .
\end{align}
The details of the derivation of the explicit form will be summarized in Appendix~\ref{App: A}, and we here show only the final expression. After some manipulation, one finds that the explicit form of $B_{\ell_1\ell_2\ell_3}$ takes the following form,
\begin{align}
B_{\ell_1\ell_2\ell_3}&=2^3\sum_{\substack{L_1,L_2,L_3}}\mathcal{I}^{2-20}_{\ell_12L_1}\mathcal{I}^{2-20}_{\ell_22L_2}\mathcal{I}^{2-20}_{\ell_32L_3}\mathcal{I}^{000}_{L_1L_2L_3}
\begin{Bmatrix}
\ell_1 & \ell_2 & \ell_3 \\
2 & 2 & 2 \\
L_1 & L_2 & L_3
\end{Bmatrix}\notag\\
&\quad\times\int x^2\D x\prod_{j=1}^3\biggl[\frac{2}{\pi}i^{L_j-\ell_j}\int\D k_j\mathcal{T}_{{\ell}_j}(k_j)j_{L_j}(k_jx)\biggr](2\pi)^4\mathcal{P}_h^2\mathcal{K}(k_1,k_2,k_3) \, , \label{eq: B-mode}
\end{align}
where 
\begin{align}
\mathcal{I}^{s_1s_2s_3}_{\ell_1\ell_2\ell_3}:=\sqrt{\frac{(2\ell_1+1)(2\ell_2+1)(2\ell_3+1)}{4\pi}}
\begin{pmatrix}
\ell_1 & \ell_2 & \ell_3 \\
s_1 & s_2 & s_3
\end{pmatrix},
\end{align}
and we imposed ${\ell}_i+L_i={\rm odd}$ $(i=1,2,3)$. We also introduced the following function,
\begin{align}
\mathcal{K}(k_1,k_2,k_3):=-\frac{(8\pi)^{3/2}}{10}\sqrt{\frac{7}{3}}f_{\rm NL}\frac{\sum_i k_i^3}{k_1k_2k_3}
\end{align}
with a dimensionless parameter $f_{\rm NL}$ defined by
\begin{align}
f_{\rm NL}:=\frac{3}{16}\frac{1-n}{n}\frac{\mu H}{\mathcal{G}_T} \, .
\end{align}
In the following section, we will show the results of the numerical calculations.

\subsection{Numerical Results}\label{Sec: results}

\begin{figure}[htb]
\begin{center}
\includegraphics[width=80mm]{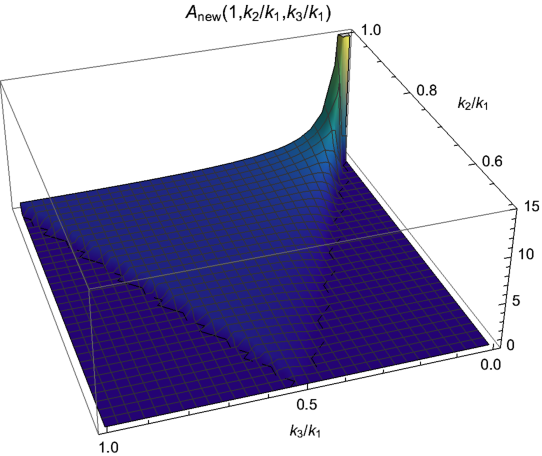}
\captionsetup{justification=raggedright}
\caption{$\mathcal{A}_{\rm new}(k_1,k_2,k_3)$ as a function of $k_2/k_1$ and $k_3/k_1$ in the case of $s_1=s_2=s_3=+2$. We have normalized the result to $1$ for the equilateral triangle $k_1=k_2=k_3$.}\label{Fig: Anew}
\end{center}
\end{figure}

\begin{figure}[htb]
\begin{center}
\includegraphics[width=80mm]{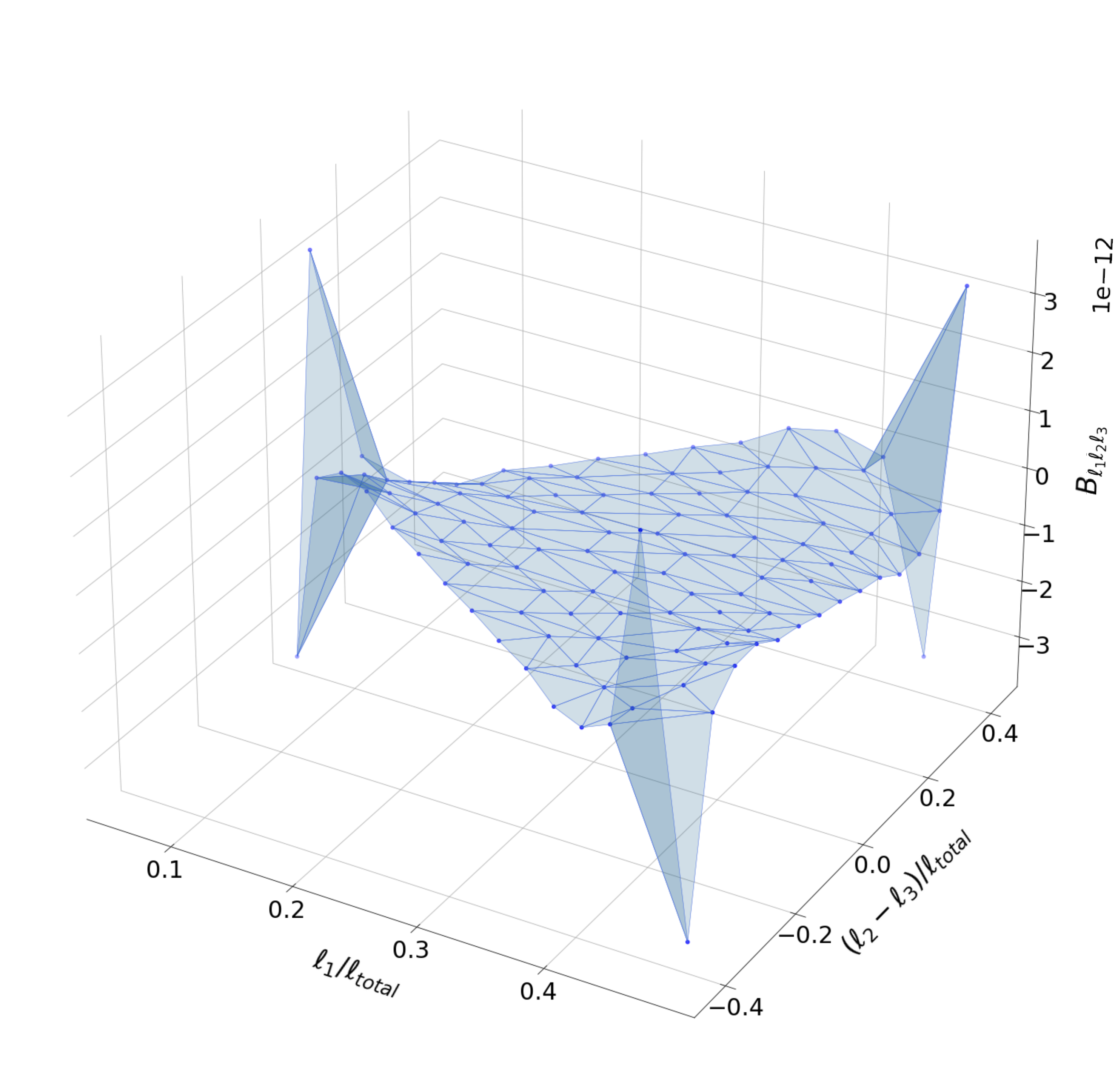}
    \captionsetup{justification=raggedright}
    \caption{The B-mode auto-bispectrum divided by $i\mathcal{P}_h^2f_{\rm NL}$ for $\ell_{\rm total}:=\sum_i \ell_i=33$. }\label{Fig: 3dplot}
\end{center}
\end{figure}

In this subsection, we will show our numerical results. The plot of the B-mode auto-bispectrum is shown in Figure~\ref{Fig: 3dplot}. For $\ell_{\rm total}=33$, the bispectrum takes the maximum value for $(\ell_1,\ell_2,\ell_3)=(2,15,16)$, similarly to the results in Refs.~\cite{Tahara:2017wud} and~\cite{Kothari:2019yyw}. Additionally, Figure~\ref{Fig: 3dplot} shows that the B-mode bispectrum has a peak at $\ell_j\ll \ell_{j+1}\simeq \ell_{j+2}$ for a given $\ell_{\rm max}$, similarly to the primordial bispectrum in Figure~\ref{Fig: Anew}, which has a peak at $k_j\ll k_{j+1}\simeq k_{j+2}$.

We then compute a signal-to-noise ratio (denoted by ${\rm SNR}$) by assuming an ideal case where the noise comes solely from cosmic variance. 
For a given maximum multipole $\ell_{\rm max}$, 
the signal-to-noise ratio squared can be computed as~\cite{Komatsu:2003iq,Shiraishi:2019yux}
\begin{align}
({\rm SNR})^2=\sum_{2\leq \ell_1\leq \ell_2\leq \ell_3\leq \ell_{\rm max}}(-1)^{\ell_1+\ell_2+\ell_3}\frac{B_{\ell_1\ell_2\ell_3}^2}{\sigma_{\ell_1\ell_2\ell_3}} \, , \label{eq: SNR}
\end{align}
where $\sigma_{\ell_1\ell_2\ell_3}$ is the variance of the bispectrum which reads~\cite{Gangui:1999vg}
\begin{align}
\sigma_{\ell_1\ell_2\ell_3}=C_{\ell_1}C_{\ell_2}C_{\ell_3}(1+\delta_{\ell_1\ell_2}+\delta_{\ell_1\ell_3}+\delta_{\ell_3\ell_2}+2\delta_{\ell_1\ell_2}\delta_{\ell_2\ell_3})
\end{align}
in an weakly-Gaussian regime where the contribution of the bispectrum to the variance can be ignored. The power spectrum of the B-mode polarizations $C_{\ell}$ is defined by
\begin{align}
\langle a^{\rm B}_{{\ell}m}a^{B}_{{\ell}'m'}\rangle=\delta_{{\ell}{\ell}'}\delta_{mm'}C_{\ell} \, ,
\end{align}
where
\begin{align}
C_l:=4\pi\int\frac{\D k}{k}\mathcal{P}_h[\mathcal{T}^B_l(k)]^2.
\end{align}

Note that the factor $(-1)^{\ell_1+\ell_2+\ell_3}$ in Eq.~(\ref{eq: SNR}) always gives $-1$ because of the odd parity property of the B-mode bispectrum (i.e., $B_{\ell_1\ell_2\ell_3}\neq0$ only for $\ell_1+\ell_2+\ell_3={\rm odd}$).

\begin{figure}[htb]
\begin{center}
\includegraphics[width=80mm]{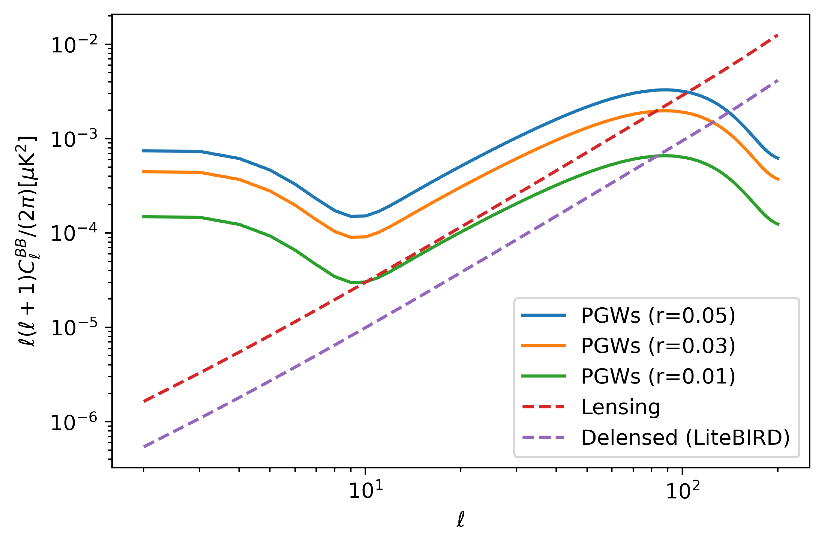}
\captionsetup{justification=raggedright}
\caption{{{The plot of the B-mode power spectrum originating from the primordial gravitational waves (PGWs) for the cases of $r=0.01, 0.03, 0.05$ (green, orange, and blue lines, respectively) and the gravitational lensing (Lensing) and the best-delensed B-mode power spectrum (Delensed) for the case of the LiteBIRD experiment. }}}\label{Fig: primlensdelens}
\end{center}
\end{figure}

\begin{figure}[htb]
\begin{center}
\includegraphics[width=80mm]{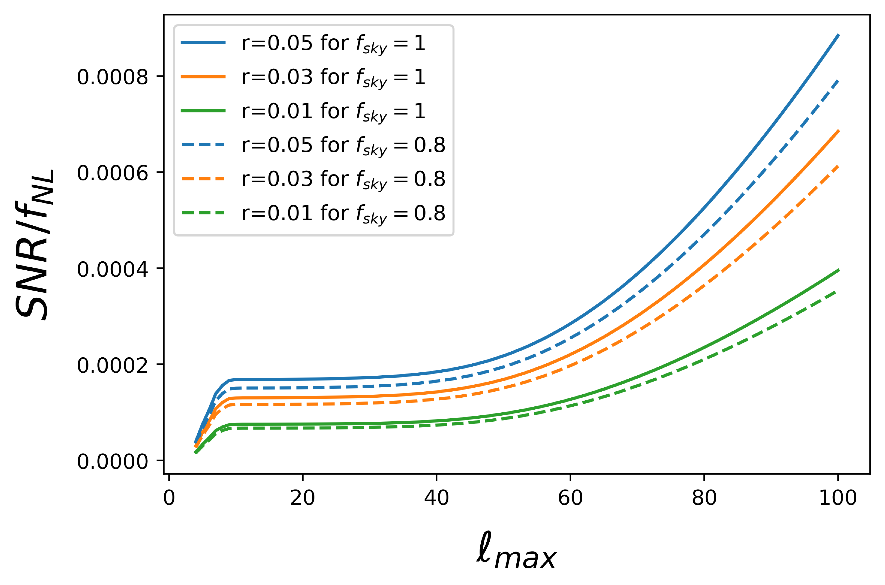}
\captionsetup{justification=raggedright}
\caption{{{The $\ell_{\rm max}$ dependence of the signal-to-noise ratio divided by $f_{\rm NL}$. The solid and dashed lines correspond to the cases of $f_{\rm sky}=1$ (full sky) and $f_{\rm sky}=0.8$ (sky coverage for the LiteBIRD experiment~\cite{Diego-Palazuelos:2020lme}), respectively, and the signal-to-noise was multiplied by $\sqrt{f_{\rm sky}}$. The cases $r=0.01, 0.03, 0.05$ correspond to the green, orange, and blue lines, respectively.} }}\label{Fig: plot-SNR}
\end{center}
\end{figure}

Here, the B-mode polarizations originate from not only the primordial gravitational waves but also the gravitational lensing~\cite{Lewis:2006fu}, and hence the lensing B-mode also acts as noise. So far, how much the lensing B-mode can be delensed has been discussed for several experiments by evaluating $\mathcal{D}:=C_l^{\rm delens}/C_l^{\rm lens}$ where $C_l^{\rm lens}$ and $C_l^{\rm delens}$ stand for a B-mode power spectrum before and after a part of the lensing effect is removed, respectively. For instance, the best possible delensing for the LiteBIRD experiment is $\mathcal{D_{\rm best}}=0.33$~\cite{Diego-Palazuelos:2020lme}. Note that, in other experiments, e.g. PICO, a smaller $\mathcal{D}_{\rm best}$ has been reported~\cite{Diego-Palazuelos:2020lme}. The B-mode power spectrum originating from the primordial gravitational waves 
 and the gravitational lensing and the best-delensed B-mode power spectrum for the case of the Lite-BIRD experiment are shown in Figure~\ref{Fig: primlensdelens}. For different values of the tensor-to-scalar ratio\footnote{The tensor-to-scalar ratio is defined as $r:=\mathcal{P}_h/\mathcal{P}_\zeta$, where $\mathcal{P}_\zeta$ is the primordial power spectrum of curvature perturbations.}, $r=0.01, 0.03, 0.05$ (below the current constraint established by CMB observations, $r<0.056$~\cite{Planck:2018jri}), the best-delensed B-mode power spectrum starts to dominate over the primordial B-mode at $\ell\sim100$, which implies that the signal-to-noise ratio would be saturated for a larger $\ell$. In light of this, let us compute the signal-to-noise ratio in the noiseless ideal case up to $\ell_{\rm max}=100$. This gives us a reliable estimation of the maximum SNR achievable with an experiment like LiteBird. The dependence of the signal-to-noise ratio on the maximum multipole $\ell_{\rm max}$ is shown in Figure~\ref{Fig: plot-SNR}.

\begin{figure}[htb]
\begin{center}
\includegraphics[width=80mm]{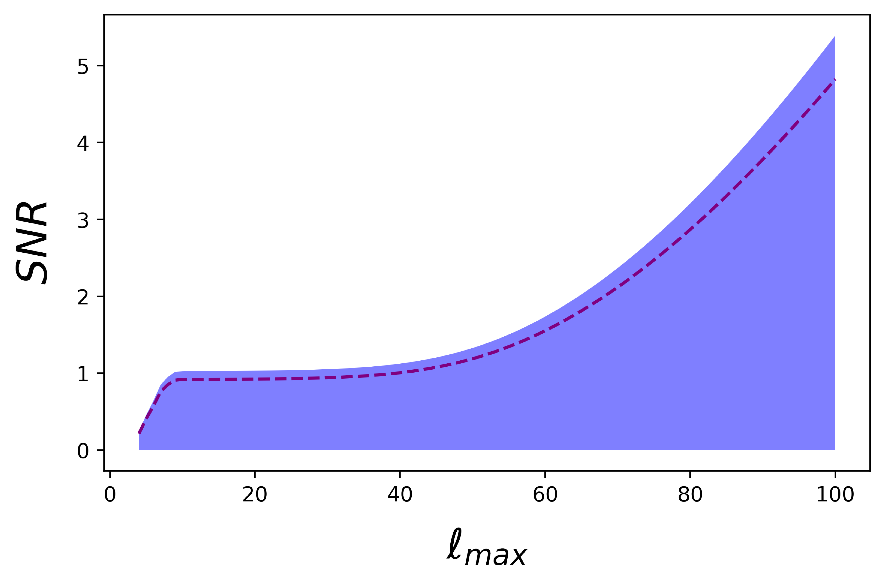}
\captionsetup{justification=raggedright}
\caption{{{The signal-to-noise ratio for the largest possible value of $f_{\rm NL}$ in the linear perturbation regime. The blue region stands for the possible values of the signal-to-noise ratio. The dashed line denotes the upper bound on the signal-to-noise ratio in light of the sky coverage for the LiteBIRD experiment, i.e., the upper bound on $\rm SNR$ multiplied by $\sqrt{f_{\rm sky}}$ where $f_{\rm sky}=0.8$~\cite{Diego-Palazuelos:2020lme}.}}}\label{Fig: plot-SNRmax}
\end{center}
\end{figure}

A larger $f_{\rm NL}$ leads to a larger signal-to-noise ratio, but the magnitude of $f_{\rm NL}$ is constrained as long as we work in a linear perturbation regime, since a larger $f_{\rm NL}$ would imply the necessity to take into account non-linear interactions (e.g. originating from the cubic self interactions of the tensor perturbations) in the equation of motion for the tensor perturbations. To discuss the validity of the predictions within the linear perturbation regime, let us evaluate the ratio of the cubic Lagrangian of the tensor perturbations (denoted by $\mathcal{L}^{(3)}_h$) to the quadratic one (denoted by $\mathcal{L}^{(2)}_h$),
\begin{align}
\frac{\mathcal{L}_h^{(3)}}{\mathcal{L}_h^{(2)}}=\frac{\mu H}{\mathcal{G}_T}\frac{|\dot h_{ij}|}{H} \, . \label{eq: cubic-to-quadratic}
\end{align}
Since $\mu H/{\mathcal{G}_T}$ is constant and $|\dot h_{ij}|/H$ monotonically increases with the conformal time, the above takes the maximum values at the end of the contracting phase at which the perturbations are on superhorizon scales. On these scales, the mode function satisfies
\begin{align}
\frac{\dot h^{(s)}_{\bf k}}{H h^{(s)}_{\bf k}}\simeq-\frac{3(1-n)}{n} \, ,
\end{align}
and hence we obtain
\begin{align}
\frac{\mathcal{L}_h^{(3)}}{\mathcal{L}_h^{(2)}}\simeq 3\frac{\mu H}{\mathcal{G}_T}\frac{1-n}{n}|h_{ij}|\simeq3\frac{\mu H}{\mathcal{G}_T}\frac{1-n}{n}\mathcal{P}_\zeta^{1/2}r^{1/2} \, , \label{eq: cubic-to-quadratic2}
\end{align}
where we used $|h_{ij}|\simeq\mathcal{P}_h^{1/2}$.\footnote{The magnitude of $h_{ij}$ is determined in terms of Fourier transformation as
$|h_{ij}|\sim \sqrt{\int\D k\mathcal{P}_h/k}$.
In our case, the tensor power spectrum is scale invariant, and we ignored the square root of the logarithmic factor, which resulted in $|h_{ij}|\simeq \mathcal{P}_h^{1/2}$.
} 
Since we evaluate Eq.~(\ref{eq: cubic-to-quadratic}) and Eq.~(\ref{eq: cubic-to-quadratic2}) at the end of the contracting phase during which the observational predictions are determined, we impose the current constraints on $\mathcal{P}_\zeta$ and $r$ as $\mathcal{P}_\zeta\simeq 2\times 10^{-9}$ and $r\leq\mathcal{O}(10^{-2})$~\cite{Planck:2018jri}. 
By assuming $\mathcal{L}^{(2)}_h>\mathcal{L}^{(3)}_h$ for the validity of the perturbative expansion\footnote{The condition $\mathcal{L}^{(2)}_h>\mathcal{L}^{(3)}_h$ naively corresponds to the one for classical non-linear corrections to the linear tensor perturbation to be small: $\mathcal{O}(|h_{ij}|f_{\rm NL})<1$ (see, e.g. Refs.~\cite{Leblond:2008gg,Baumann:2011dt,Joyce:2011kh} for related discussions for scalar non-Gaussianities in the context of inflation). Additionally, one might expect that strong coupling could appear on the subhorizon scales. For the canonically normalized perturbations, one obtains the quadratic and cubic actions as
\begin{align}
S^{(2)}_v&=\frac{1}{2}\int\D y\D^3x \biggl[\biggl(\frac{\D v_{ij}}{\D y}\biggr)^2-(\partial_k v_{ij})^2+\frac{1}{z}\frac{\D^2 z}{\D y^2}v_{ij}^2\biggr],\\
S^{(3)}_v&=\int\D y\D^3x \biggl[\frac{1}{z}\biggl(v_{ik}v_{jl}-\frac{1}{2}v_{ij}v_{kl}\biggr)v_{ij,kl}+\frac{a\mu c_h^2}{12}\cdot\frac{\D}{\D y}\biggl(\frac{v_{ij}}{z}\biggr)\cdot\frac{\D}{\D y}\biggl(\frac{v_{jk}}{z}\biggr)\cdot\frac{\D}{\D y}\biggl(\frac{v_{ki}}{z}\biggr)\biggr] \, ,
\end{align} 
where $v_{ij}=zh_{ij}, z:=a(\mathcal{G}_T\mathcal{F}_T)^{1/4}/2$, and $\D y:=c_h\D t/a$. In particular, the second term in the cubic action scales on subhorizon scales ($-ky\gg1$) as
\begin{align}
\int\D y\D^3x \biggl[\frac{a\mu c_h^2}{12}\cdot\frac{\D}{\D y}\biggl(\frac{v_{ij}}{z}\biggr)\cdot\frac{\D}{\D y}\biggl(\frac{v_{jk}}{z}\biggr)\cdot\frac{\D}{\D y}\biggl(\frac{v_{ki}}{z}\biggr)\biggr]\sim \int\D y\D^3x \frac{1}{\Lambda^2}\cdot\frac{\D v_{ij}}{\D y}\cdot\frac{\D v_{jk}}{\D y}\cdot\frac{\D v_{ki}}{\D y} \, , 
\end{align}
where $\Lambda\propto (-y)^{1/2}$ which is asymptotic to $\infty$ for $y\to -\infty$. Therefore, the $\dot h_{ij}^3$ term is not dangerous and strong coupling does not appear on the subhorizon scales. See also Refs.~\cite{Ageeva:2018lko,Ageeva:2020buc,Ageeva:2020gti,Ageeva:2021yik,Ageeva:2022asq,Ageeva:2022fyq,Akama:2022usl,Akama:2024bav} for related discussions where coupling functions of cubic scalar and/or tensor interactions are asymptotic to infinity ($\Lambda\to0$ in the above case) in the past infinity.}, we obtain
\begin{align} 
\frac{\mu H}{\mathcal{G}_T}\frac{1-n}{n}<\frac{1}{3}\mathcal{P}_\zeta^{-1/2}r^{-1/2} \, .
\end{align}
Finally, the constraint on $f_{\rm NL}$ reads
\begin{align}
f_{\rm NL}=\frac{3}{16}\frac{1-n}{n}\frac{\mu H}{\mathcal{G}_T}< \frac{1}{16}\mathcal{P}_\zeta^{-1/2}r^{-1/2} \, .
\end{align}
Here, ${\rm SNR}/f_{\rm NL}$ and the upper bound on $f_{\rm NL}$ are proportional to $r^{1/2}$ and $r^{-1/2}$, respectively. Therefore, the upper bound on the signal-to-noise ratio is independent of the value of $r$. A plot of the resultant possible values of the signal-to-noise ratio is given in Figure~\ref{Fig: plot-SNRmax}. We find that there is a parameter space in which the signal-to-noise ratio can reach unity for $\ell_{\rm max}\geq9$. In particular, the possible maximum value is about $5.39$ for $\ell_{\rm max}=100$. When we take into account the sky coverage for the LiteBIRD experiment, the signal-to-noise ratio can reach unity for $\ell_{\rm max}=40$ and its possible maximum value is about $4.82$ for $\ell_{\rm max}=100$.

Before closing this section, let us comment on the expected B-mode signals from one of the non-standard models of inflation that involves a non-attractor phase making the tensor perturbations grow in time as in a matter-dominated contracting universe~\cite{Ozsoy:2019slf}. The dynamics of the perturbations in that paper is identical to ours. Actually, as has been studied in~\cite{Ozsoy:2019slf}, the three-point function of the tensor perturbations from that non-attractor model of inflation have the same momentum dependence with that from the framework in our paper. Therefore, our results indicate that the non-attractor model of inflation also has the potential to be tested with the B-mode auto-bispectrum by the upcoming CMB experiments.

\section{Summary}\label{Sec: conclusion}
In this paper, we have investigated the three-point statistics of the B-mode polarization in the general bounce cosmology. In particular, by focusing on the newly induced cubic operator of the tensor perturbations in the Horndeski theory, we have computed the B-mode auto-bispectrum and its signal-to-noise ratio for the noiseless case. In doing so, we have evaluated the theoretical upper bound on the signal-to-noise ratio by assuming the linear perturbation regime and showed that the signal-to-noise ratio can reach unity for $\ell_{\rm max}\geq9$ and $\ell_{\rm max}\geq40$ in the full sky and the sky coverage for the LiteBIRD experiment, respectively, and can increase up to $5.39$ and $4.82$ in the former and the later, respectively, for $\ell_{\rm max}=100$. In the context of the non-singular alternatives, we have shown the first example which can predict ${\rm SNR}\geq1$.
Once the B-mode bispectrum is detected, the extended bounce models can be one of the candidates successfully explaining the origin of the signal, in addition to inflationary models. Therefore, this project has its importance also from the viewpoint of testing the inflationary paradigm in terms of its counterpart.

In addition to the results presented in this paper, it would be interesting to take into account experimental and lensing B-mode noise to discuss the possibility of the detection of the B-mode bispectrum more concretely, as has been done in the context of inflation in Ref.~\cite{Shiraishi:2019yux}. Additionally, only a few examples predicting sizable three-point statistics of the B-mode polarization have been found even in the context of inflation, and thus it would still be interesting to find early universe scenarios predicting enhanced tensor non-Gaussianities and a sizable B-mode bispectrum. Since the amplitude of the curvature perturbation is larger than that of gravitational waves, it would also be relevant to investigate the CMB cross bispectra originating from primordial cross bispectra in the general bounce cosmology. We leave these investigations to our future work.

\section*{Acknowledgements}
We thank Takashi Hiramatsu, Chunshan Lin, and Hiroaki Tahara for useful discussions. G.O. thanks the National Energy Research Scientific Computing Center for providing access to the Perlmutter computing cluster. P.C.M.D. thanks Ruth Durrer and the University of Geneva for providing access to the Baobab cluster. The work of S.A. was supported by the grant No. UMO-2021/42/E/ST9/00260 from the National Science Centre, Poland, and MEXT-JSPS Grant-in-Aid for Transformative Research Areas (A) ``Extreme Universe'', No.~JP21H05189. The work of G.O. was supported by the grant No. UMO-2021/42/E/ST9/00260 from the National Science Centre, Poland. The work of P.C.M.D was supported by the grant No. UMO-2021/42/E/ST9/00260 from the National Science Centre, Poland.

\appendix
\section{Derivation of the B-mode Bispectrum}\label{App: A}
One can expand the delta function and the polarization tensor in terms of the (spin-weighted) spherical harmonics~\cite{Shiraishi:2010kd}
\begin{align}
\delta^{(3)}({\bf k}_1+{\bf k}_2+{\bf k}_3)&=8\int_0^{\infty}x^2\D x\prod_{n=1}^3\sum_{L_n,M_n}(-1)^{L_n/2}j_{L_n}(k_n x)Y^*_{L_nM_n}(\hat n_n)\mathcal{I}^{000}_{L_1L_2L_3}
\begin{pmatrix}
L_1 & L_2 & L_3 \\
M_1 & M_2 & M_3
\end{pmatrix},\\
e^{(s)}_{ij}(\hat k)&=\frac{3}{\sqrt{2\pi}}\sum_{M,m_i,m_j}{}_{-s}Y^*_{2M}(\hat k)\alpha^{m_i}_i\alpha^{m_j}_j
\begin{pmatrix}
2 & 1 & 1 \\
M & m_i & m_j
\end{pmatrix} \, , \label{eq: polarization}
\end{align}
where
\begin{align}
(\alpha^m)_a&:=\sqrt{\frac{2\pi}{3}} \label{eq: def-alpha-vec}
\begin{pmatrix}
 -m(\delta_{m,1}+\delta_{m,-1}) \\ i(\delta_{m,1}+\delta_{m,-1}) \\ \sqrt{2}\delta_{m,0} 
\end{pmatrix} \, .
\end{align}
By using Eqs.~({\ref{eq: polarization}}) and~(\ref{eq: def-alpha-vec}), we obtain
\begin{align}
e^{*(s_1)}_{ij}(\hat k_1)e^{*(s_2)}_{jk}(\hat k_2)e^{*(s_3)}_{ki}(\hat k_3)=-(8\pi)^{3/2}\frac{1}{10}\sqrt{\frac{7}{3}}\sum_{M,M',M''}{}_{s_1}Y^*_{2M}(\hat k_1){}_{s_2}Y^*_{2M'}(\hat k_2){}_{s_3}Y^*_{2M''}(\hat k_3)
\begin{pmatrix}
2 & 2 & 2 \\
M & M' & M''
\end{pmatrix} \, ,
\end{align}
where we also used the following property of the Wigner symbols:
\begin{align}
&\sum_{m_i,m_j,m_k}(-1)^{3-m_i-m_j-m_k}
\begin{pmatrix}
1 & 2 & 1 \\
m_i & -M & m_j
\end{pmatrix}
\begin{pmatrix}
1 & 2 & 1 \\
m_j & -M' & m_k
\end{pmatrix}
\begin{pmatrix}
1 & 2 & 1 \\
m_k & -M'' & m_i
\end{pmatrix}\notag\\
\quad &=
\begin{pmatrix}
2 & 2 & 2 \\
M & M' & M''
\end{pmatrix}
\begin{Bmatrix}
2 & 2 & 2 \\
1 & 1 & 1
\end{Bmatrix}\notag\\
\quad &=\frac{1}{10}\sqrt{\frac{7}{3}}
\begin{pmatrix}
2 & 2 & 2 \\
M & M' & M''
\end{pmatrix} \, ,
\end{align}
and 
$\begin{Bmatrix}
\ell_1 & \ell_2 & \ell_3 \\
m_1 & m_2 & m_3
\end{Bmatrix}$
denotes the Wigner-$6j$ symbol.

Then, $B^{(s_1s_2s_3)}_{\ell_1\ell_2\ell_3}$ is obtained as
\begin{align}
B^{(s_1s_2s_3)}_{\ell_1\ell_2\ell_3}&=\sum_{\substack{L_1,L_2,L_3}}\mathcal{I}^{s_1-s_10}_{\ell_12L_1}\mathcal{I}^{s_2-s_20}_{\ell_22L_2}\mathcal{I}^{s_3-s_30}_{\ell_32L_3}\mathcal{I}^{000}_{L_1L_2L_3}
\begin{Bmatrix}
\ell_1 & \ell_2 & \ell_3 \\
2 & 2 & 2 \\
L_1 & L_2 & L_3
\end{Bmatrix}\notag\\
&\quad\times\int x^2\D x\prod_{j=1}^3\biggl[\frac{2}{\pi}\biggl(\frac{s_j}{2}\biggr)i^{L_j-\ell_j}\int\D k_j\mathcal{T}_{\ell_j}(k_j)j_{L_j}(k_jx)\biggr](2\pi)^4\mathcal{P}_h^2\mathcal{K}(k_1,k_2,k_3) \, , \label{eq: B-mode}
\end{align}
where 
$\begin{Bmatrix}
\ell_1 & \ell_2 & \ell_3 \\
m_1 & m_2 & m_3 \\
L_1 & L_2 & L_3
\end{Bmatrix}
$ is the Wigner-$9j$ symbol, and we used the following properties of the spin-weighted spherical harmonics and the Wigner symbols:
\begin{align}
&\int \D \Omega_k{}_{-s_j}Y^*_{\ell_jm_j}(\Omega_{k_j})Y^*_{L_nM_n}(\Omega_{k_j}){}_{s_j}Y^*_{2M}(\Omega_{k_j})=\mathcal{I}^{s_j 0 -s_j}_{\ell_j L_j 2}
\begin{pmatrix}
\ell_j & L_j & 2 \\
m_j & M_j & M
\end{pmatrix},\\
&\sum_{\substack{ M_1,M_2,M_3 \\ M,M',M''}}
\begin{pmatrix}
L_1 & L_2 & L_3 \\
M_1 & M_2 & M_3
\end{pmatrix}
\begin{pmatrix}
2 & 2 & 2 \\
M & M' & M''
\end{pmatrix}
\begin{pmatrix}
\ell_1 & L_1 & 2 \\
m_1 & M_1 & M
\end{pmatrix}
\begin{pmatrix}
\ell_2 & L_2 & 2 \\
m_2 & M_2 & M'
\end{pmatrix}
\begin{pmatrix}
\ell_3 & L_3 & 2 \\
m_3 & M_3 & M''
\end{pmatrix}\notag\\
&\quad =
\begin{pmatrix}
\ell_1 & \ell_2 & \ell_3 \\
m_1 & m_2 & m_3
\end{pmatrix}
\begin{Bmatrix}
\ell_1 & \ell_2 & \ell_3 \\
L_1 & L_2 & L_3 \\
2 & 2 & 2
\end{Bmatrix},\\
&\sum_{m_1,m_2,m_3}
\begin{pmatrix}
\ell_1 & \ell_2 & \ell_3 \\
m_1 & m_2 & m_3
\end{pmatrix}
\begin{pmatrix}
\ell_1 & \ell_2 & \ell_3 \\
m_1 & m_2 & m_3
\end{pmatrix}
=1 \, .
\end{align}
By taking the sum over the helicities, we obtain
\begin{align}
B_{\ell_1\ell_2\ell_3}&=\sum_{s_1,s_2,s_3}B^{(s_1s_2s_3)}_{\ell_1\ell_2\ell_3}\notag\\
&=2^3\sum_{\substack{L_1,L_2,L_3}}\mathcal{I}^{2-20}_{\ell_12L_1}\mathcal{I}^{2-20}_{\ell_22L_2}\mathcal{I}^{2-20}_{\ell_32L_3}\mathcal{I}^{000}_{L_1L_2L_3}
\begin{Bmatrix}
\ell_1 & \ell_2 & \ell_3 \\
2 & 2 & 2 \\
L_1 & L_2 & L_3
\end{Bmatrix}\notag\\
&\quad\times\int x^2\D x\prod_{j=1}^3\biggl[\frac{2}{\pi}i^{L_j-\ell_j}\int\D k_j\mathcal{T}_{\ell_j}(k_j)j_{L_j}(k_jx)\biggr](2\pi)^4\mathcal{P}_h^2\mathcal{K}(k_1,k_2,k_3) \, , \label{eq: B-mode}
\end{align}
with $\ell_i+L_i={\rm odd}$ $(i=1,2,3)$ being imposed. To derive the above expression, we used
\begin{align}
\sum_{s}\frac{s}{2}\mathcal{I}^{s0-s}_{\ell L2}&=[1-(-1)^{\ell+L}]\mathcal{I}^{20-2}_{\ell L2}\notag\\
&=\begin{cases}
\displaystyle{2}\mathcal{I}^{20-2}_{\ell L2}
 &\ \ (\ell+L={\rm odd}) \, ,\\
\displaystyle{0}
 &\ \ (\ell+L={\rm even}) \, .
\end{cases}
\end{align}


\bibliography{Bmode}

\end{document}